\documentclass{acm_proc_article-sp}

\usepackage[utf8]{inputenc}
\usepackage[usenames]{color}
\usepackage{graphicx}
\usepackage[pdftex, 
  linktocpage=true,
  hypertexnames=true, 
  colorlinks=true, 
  citecolor=blue,
  bookmarks=true, 
  bookmarksopen=true, 
  pdftitle=Towards\ Bridging\ IoT\ and\ Cloud\ Services:\ Enabling\ Smartphones\ as\ Mobile\ and\ Autonomic\ Service Gateways, 
  pdfauthor=Roya Golchay\ -\ Frédéric\ Le\ Mouël\ -\ Stéphane\ Frénot\ -\ Julien\ Ponge, 
  pdfsubject=Position\ Paper, 
  pdfcreator=dvipdf,
  pdfkeywords=Cloud\ Computing\ -\ Internet\ of\ Things\ -\ Service-oriented\ Applications\ -\ Middleware\ -\ Smart\ and\ Autonomic\ Gateways\ -\ Smartphone]{hyperref}

\pdfoutput=1

\DeclareGraphicsExtensions{.pdf}

\begin{document}

%\title{Towards Bridging IoT and Cloud Services: Enabling Smartphones as Mobile and Autonomic Service Gateways}
\title{Towards Bridging IoT and Cloud Services: Proposing Smartphones as Mobile and Autonomic Service Gateways}

\subtitle{Position Paper}

\numberofauthors{4} 

\author{
% You can go ahead and credit any number of authors here,
% e.g. one 'row of three' or two rows (consisting of one row of three
% and a second row of one, two or three).
%
% The command \alignauthor (no curly braces needed) should
% precede each author name, affiliation/snail-mail address and
% e-mail address. Additionally, tag each line of
% affiliation/address with \affaddr, and tag the
% e-mail address with \email.
%
% 1st. author
\alignauthor
Roya Golchay\\
       \affaddr{Université de Lyon, INRIA}\\
       \affaddr{INSA-Lyon, CITI}\\
       \affaddr{F-69621, Villeurbanne, France}\\
       \email{\mbox{roya.golchay@insa-lyon.fr}}
% 2nd. author
\alignauthor
Frédéric Le Mouël\\
       \affaddr{Université de Lyon, INRIA}\\
       \affaddr{INSA-Lyon, CITI}\\
       \affaddr{F-69621, Villeurbanne, France}\\
       \email{frederic.le-mouel@insa-lyon.fr}
% 3rd. author
\alignauthor 
Stéphane Frénot\\
       \affaddr{Université de Lyon, INRIA}\\
       \affaddr{INSA-Lyon, CITI}\\
       \affaddr{F-69621, Villeurbanne, France}\\
       \email{\mbox{stephane.frenot@insa-lyon.fr}}
\and  % use '\and' if you need 'another row' of author names
% 4th. author
Julien Ponge\\
       \affaddr{Université de Lyon, INRIA}\\
       \affaddr{INSA-Lyon, CITI}\\
       \affaddr{F-69621, Villeurbanne, France}\\
       \email{\mbox{julien.ponge@insa-lyon.fr}}
}
% There's nothing stopping you putting the seventh, eighth, etc.
% author on the opening page (as the 'third row') but we ask,
% for aesthetic reasons that you place these 'additional authors'
% in the \additional authors block, viz.
%\additionalauthors{}
% Just remember to make sure that the TOTAL number of authors
% is the number that will appear on the first page PLUS the
% number that will appear in the \additionalauthors section.

%\date{30 July 1999}

\maketitle

\begin{abstract}
Computing is currently getting at the same time incredibly in the small with sensors/actuators embedded in our everyday objects and also greatly in the large with data and service clouds accessible anytime, anywhere. This Internet of Things is physically closed to the user but suffers from weak run-time execution environments. Cloud Environments provide powerful data storage and computing power but can not be easily accessed and integrate the final-user context-awareness. We consider smartphones are set to become the universal interface between these two worlds. In this position paper, we propose a middleware approach where smartphones provide service gateways to bridge the gap between IoT services and Cloud services. Since smartphones are mobile gateways, they should be able to (re)configure themself according to their place, things discovered around, and their own resources such battery. Several issues are discussed: collaborative event-based context management, adaptive and opportunistic service deployment and invocation, multi-criteria (user- and performance-oriented) optimization decision algorithm.

%As computers get smaller, portables or even embeded into everyday objects, the internet becomes the internet of things and it is popular to be able to access to internet from anything, anytime and everywhere. As the mobile phones are one of the most numerous mobile computers, in this position paper we consider them as an appropriate mobile gateway to provide access to online services and cloud computing applications in more efficient way. Since this gateway is a mobile gateway it should be able to configure itself according to its place and the thing arount. On the other hand limited battery life and resources force the gateway to decide where applications should be run in order to have more efficiens services with minimum resource utilisation. In this paper we disscuss the algorithms to have an inteligent mobile gateway and its challengaes.
\end{abstract}

% Categories



\section{Introduction}

Computing is currently getting at the same time incredibly in the small with sensors/actuators embedded in our everyday objects and also greatly in the large with data and service clouds accessible anytime, anywhere. This Internet of Things is physically closed to the user but suffers from weak run-time execution environments. Cloud Environments virtualize and provide powerful data storage and computing power but can not be easily accessed and integrate the final-user context-awareness. 

In parallel with these run-time environment evolutions, the service-oriented programming and architectures are also important evolutions in software engineering. These paradigms allow to easily split applications, enable a great reuse and (sometimes) handle (some) communication protocol issues. New run-time challenges are still raised especially with IoT and Cloud environments such as service deployment mechanisms, service placement, service invocation scheduling, etc.

\begin{figure}[!htb]
\centering
\psfig{file=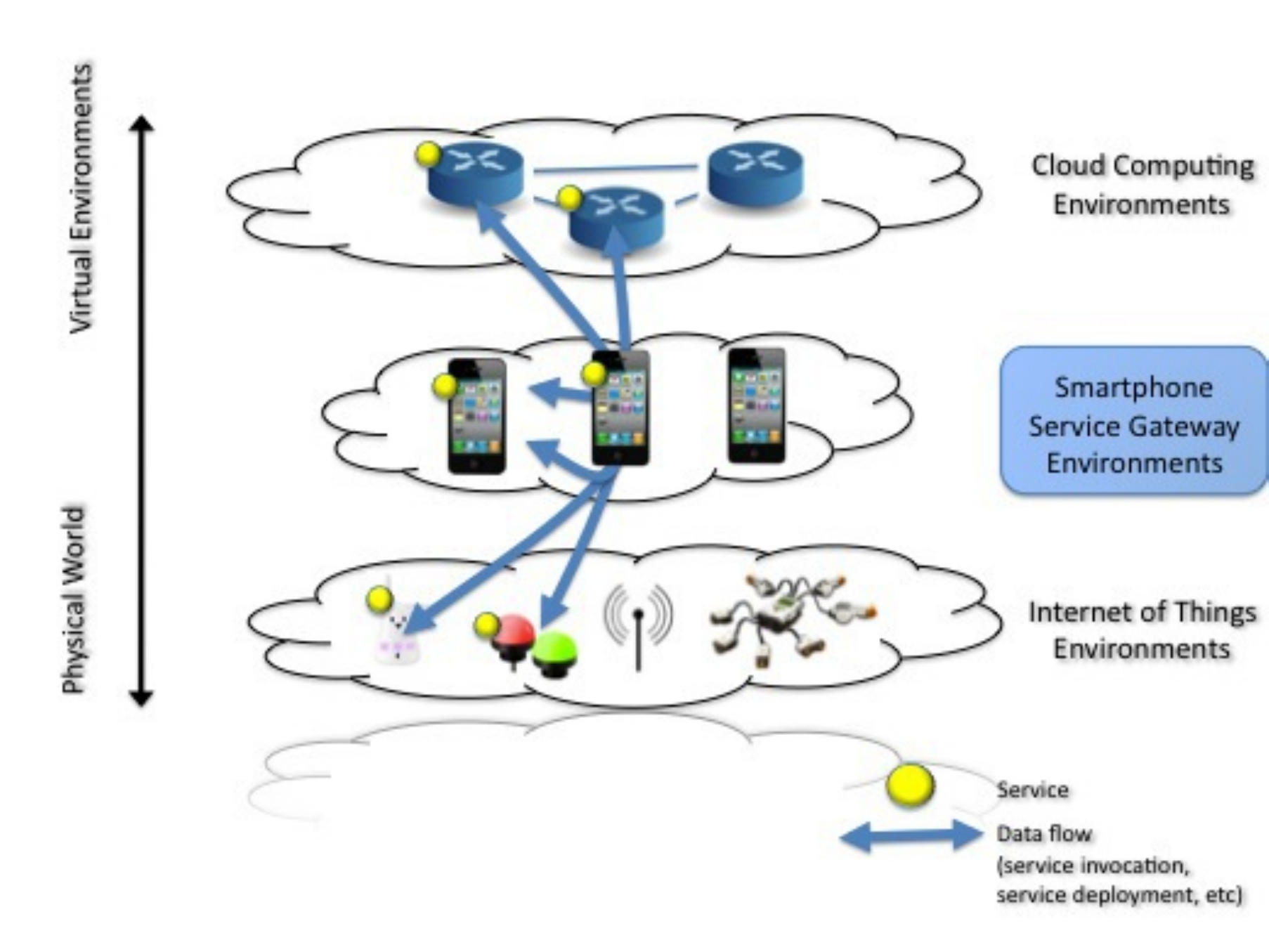, width=\columnwidth, bb=0 0 520 390}
\caption{Smartphone Service Gateways Bridging IoT and Cloud Services}
\label{fig:smartgateway}
\end{figure}

We consider smartphones are set to become the universal interface between Internet of Things and Cloud Computing worlds, as shown in Figure~\ref{fig:smartgateway}. In this position paper, we propose a middleware approach where smartphones provide service gateways to bridge the gap between IoT services and Cloud services. Since smartphones are mobile gateways, they should be able to (re)configure themself according to their place, things discovered around, and their own resources such battery. Several issues are discussed to implement a smart and autonomic service gateway: collaborative event-based context management, adaptive and opportunistic service deployment and invocation, multi-criteria (user- and performance-oriented) optimization decision algorithm.

The rest of this paper is organized as follow. Next section presents related works. An overview of our proposed solution is given in section~\ref{sec:overview} and smart and autonomic mechanisms are detailed in section~\ref{sec:mechanims}. Finally the last section states our current and future works.

\section{Related Work}
\label{sec:relatedwork}

While talking about mobile environment the first thing that attracts attention is resource-poor devices compared with fixed stations.  Mobile phones are limited in computation power, available memory and battery life. The wireless connections with different available bandwidth and probability of disconnection and reconnection are another important point in these environments. There is a considerable amount of research on how to address these mobility issues by migrating data and computation to distributed systems~\cite{LeMF:2002}. This issue is now raised by Internet of Things where things are even more constrained. To address the low energy power issue, \cite{Kumar:2010} proposed using cloud computing. According to this article sending computation to the cloud will save energy on smart phone. The amount of saved energy is depends on the wireless bandwidth, the amount of data to be transmitted to cloud and also the application type if it is real-time data application or not. In order to use the cloud more efficiently, how to distribute the application is very important. Wishbone~\cite{Newton:2009} is a system that takes a data-flow graph of operators and produces an optimal partitioning. The partitioning problem is to find a cut of data-flow graph, where its vertices are stream operators and its edges are streams. Edge and vertex weights represent bandwidth and CPU utilisation respectively. The partitioning algorithm models the program as an integer linear program that minimizes a linear combination of network bandwidth and CPU load. The notion of data-flow graph is also used in~\cite{Gi:2009} where the presented middleware platform can automatically distribute different layers of an application between the phone and the cloud. In this graph the vertexes are software modules where service dependency between them represents by the edges. The optimal cut of the graph will minimizes or maximizes the objective function that can be the end-to-end interaction time between a phone and a server, the amount of exchanged data or complete the execution in less than a predefined time. 
The layered architecture is also considered in~\cite{Gnawali:2006} where presented tenet architecture for tiered sensor networks networks simplified application development. This architecture composed of two tiers: a lower tier consisting of motes, which enable flexible deployment of dense instruction, and upper tier containing fewer, relatively less constrained nodes. 
Another three tiers management architecture is used in~\cite{Bourcier:2011} where a service-oriented framework is introduced to simplify the development and run-time adaptive support of autonomic pervasive applications. A middleware is proposed to serve as a framework to host autonomic home applications. A distributed resource model and management tailored for deployment of adaptive services in a mobile environment is also discussed in~\cite{Alia:2007}. This research is focuses on the resources directly surrounding one mobile user. The presented  distributed resource management framework is used to monitor the context and dynamically respond to changes by selecting the most appropriate application variant which called self-adaption mechanism.
Although the distributed powerful resources in the cloud seems useful to cover the resource constraints of small components in Internet of Things, absence of context-awareness in the cloud make it difficult to communicate with it for the users. Several bridging approaches exist~\cite{Bissyande:2011} but a link is required to handle communication between almost physical environment of IoT and virtual environment of the cloud.

\section{Solution Overview}
\label{sec:overview}

We consider service-oriented applications modeled as a graph where nodes are services and vertices are service dependencies. Two service dependencies are possible: static deployment dependency and dynamic run-time invocations.

\begin{figure}[!htb]
\centering
\psfig{file=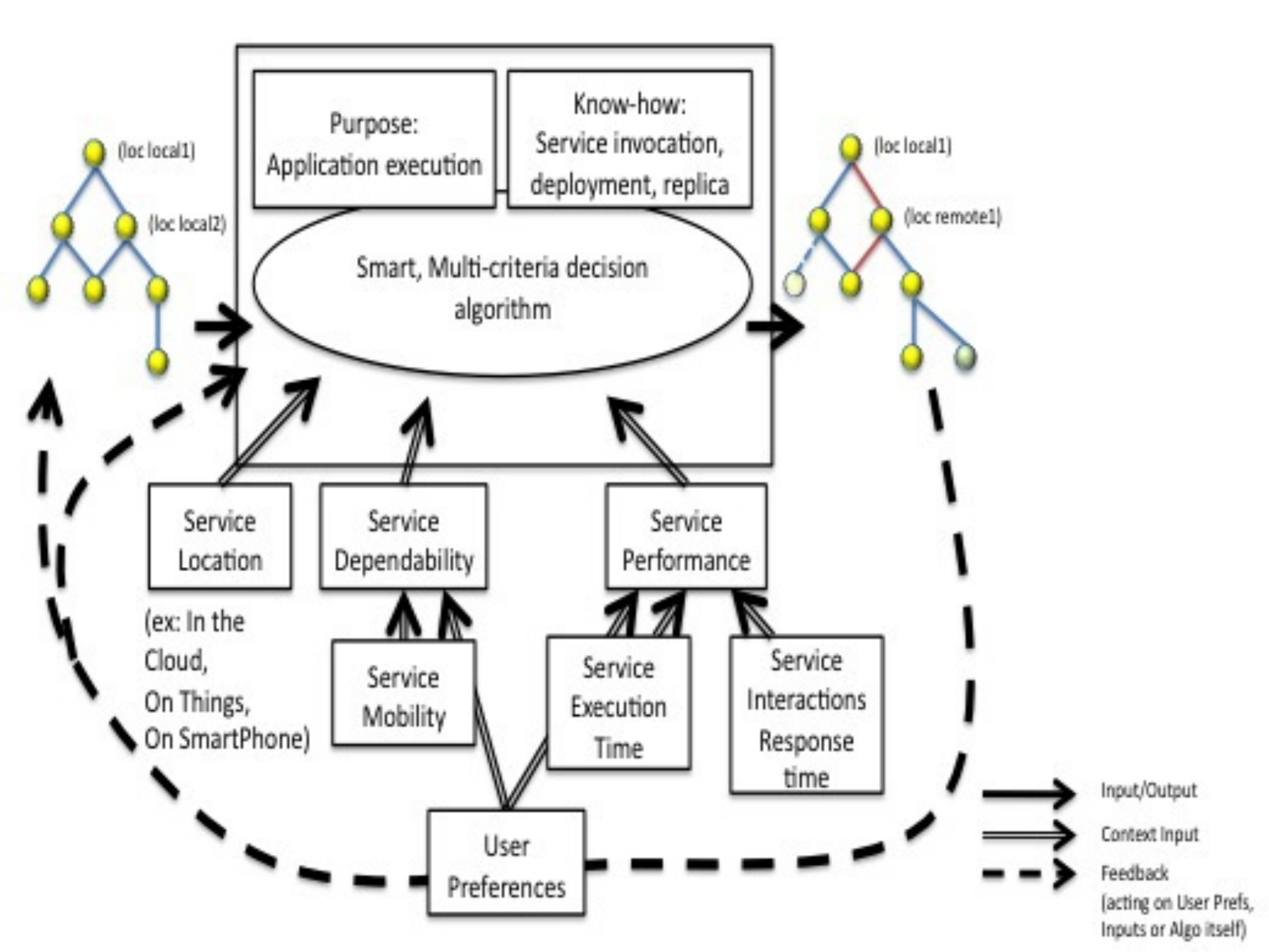, width=\columnwidth, bb=0 0 520 390}
\caption{Smartphone Service Gateway - Conceptual Model}
\label{fig:model}
\end{figure}

For the smartphone service gateway we propose, as shown in Figure~\ref{fig:smartgateway}, the major challenge consists in deploying, invoking, (and eventually replicating) services of an application by integrating (i)~existing light-services on communicating objects from IoT, (ii)~mid-services on the smartphones closer to the end-users, (iii)~heavy-services distributed in the Cloud.

The proposed conceptual model in Figure~\ref{fig:model} has this service graph in input, applies a multi-criteria decision algorithm and realizes deployment/invocation/(replication) actions to schedule/coordinate the service choreography. The decision algorithm takes into account different context inputs extracted from the services (size, performance, etc), from plateforms (mobility, location, etc) and from the user (preferences, profile, etc).

This process is autonomic because we think a smartphone has especially to be self-managed, without the user intervention. It is however important the decision algorithm to be user-oriented because a smartphone is really an end-user terminal. The process includes so a retro-action loop from service uses/monitoring feedbacks. This retro-action loop allows to automatically modify the user profile, to change the algorithm inputs by reannotating the service graph, or to adapt the decision algorithm it-self.

%It seems that using the gateway is a solution to cover the resource-poorness of embedded mini-computers by connecting them to the powerful distributed computers in the cloud. Due to the increasing number of the smartphones and their vast dispersion they are good choices to play gateway role.
%We consider a group of smartphones as gateway layer. In fact these phones are collaborate to prepare more efficient services with the minimum cost. This gateway layer should be intelligent, these smartphones should communicate to decide if they can give the services to each other, and also to decide which part of the services should be run on the cloud. While talking the mobile gateway, adaptability is another important point to consider. As soon as a smartphone move, it should be able to detect its new environment and communicate either with the existing gateways, with the IoT and cloud. It should also be able to hand over the services to other gateways while leaving. Preparing reasonable services considering the network bandwidth, battery life time and type of the service is what we expect this gateway layer.

\section{Smart and Autonomic Gateway Mechanisms}
\label{sec:mechanims}

\begin{figure}[!htb]
\centering
\psfig{file=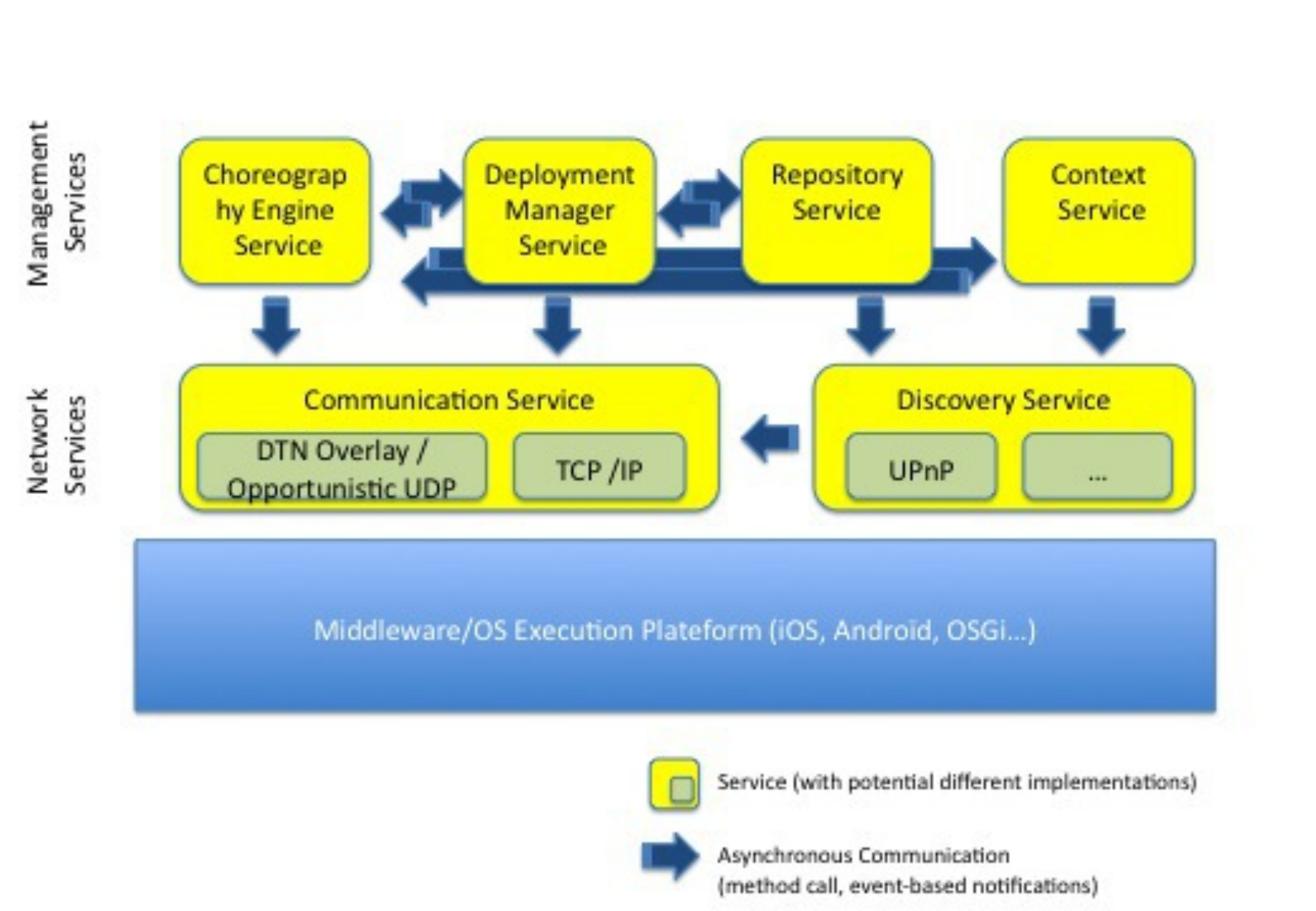, width=\columnwidth, bb=0 0 520 360}
\caption{Smartphone Service Gateway - Architecture Overview}
\label{fig:architecture}
\end{figure}

After analysing the properties of autonomic systems, we propose several mechanisms implemented in a service-oriented middleware approach (cf Figure~\ref{fig:architecture}).

\emph{Aware property}
\vspace*{-10pt}
\begin{enumerate}
\item \textbf{Event-based context management}: The context management is crucial in an autonomous system while these inputs are conditioning the decision. In our approach, we focus the context monitoring on services: their performance, location, mobility, etc. In very highly changing environments such as IoT and Cloud ones, to have a constantly reevaluating algorithm, we use a \emph{Listener/Event Notifier} approach for each changes. Pertinently qualifying the moment to send an event is critical on a smartphone because a too active monitoring leads to a battery run off and a too passive monitoring do not produce a accurate view of the physical world. We will examine context aggregation and filtering such as in~\cite{Riboni:2011}.
\item \textbf{Collaborative services repositories}: Services running on a smartphone can be available for other smartphones in vicinity. Each smartphone holds a local \emph{Repository Service} that can be discovered by using different discovery protocols. Smartphones can then collaborate, for example, to use all the same service and reduce battery consumption of participants. Aimed collaborative approaches are not those of distributed systems with strong consensus or transaction problems, either are geographically-closed and short-lived collaborations with only local stability.
\end{enumerate}

\emph{Adaptive property}
\vspace*{-10pt}
\begin{enumerate}
\setcounter{enumi}{2}
\item \textbf{Adaptive, delayed and opportunistic service deployment and invocation}: Deployment and coordination mechanisms need to be adaptive to allow migration or call redirection according to smartphones mobility and/or disconnections. We will use a DTN communication stack implementation (Disrupted-Tolerant Network) in the \emph{Communication Service} to provide on-demand deployment and most of all - spontaneous and opportunistic deployment such as in~\cite{Guidec:2010}. The opportunistic deployment can for instance be implemented by caching services prioritarily from well-known users or from users sharing common interests such as in Social Networks.
\end{enumerate}

\emph{Automatic property}
\vspace*{-10pt}
\begin{enumerate}
\setcounter{enumi}{3}
\item \textbf{Smart multi-criteria decision algorithm}: Our Service Gateway bridging several worlds - IoT, Cloud, Smartphones and End-users -, the decision algorithm implemented in the \emph{Choreography Engine Service} has to consider contextual events from the physical environment such as the service location or mobility, from the virtual environment such as the service execution performances, from plateforms such as the battery level and from user such as the service use statistics. This problem of solving multiple constraints is NP-complete but several approaches seems very interesting for mobile environments such as approximation algorithms for constrained knapsack problems~\cite{Borradaile:2010} or biologically-inspired algorithms~\cite{Hiroshi:2011,Altman:2010}. Using the feedback control loop, we will particularly enrich them with a smart learning of the user profile.

\end{enumerate}

\section{Current and Future Work}
\label{sec:futurework}

A prototype - AxSeL\footnote{\url{http://amazones.gforge.inria.fr/}, INRIA Amazones Team Forge}, A conteXtual SErvice Loader - was implemented based on a Felix\footnote{\url{http://felix.apache.org/}}, one Java implementation of the OSGi service-oriented specification\footnote{\url{http://www.osgi.org/}}~\cite{Hamida:2008,Hamida:2011}. This prototype can run on constrained devices and locally manages service-oriented application deployment according to hardware resources. It has been tested with a PDF reader application.

As immediate extensions, we will improve the prototype and focus on Smart Building domain with the INSA Lyon project: ``The Smart Chappe Building: a Plateform for Contextual Services''~\footnote{\url{http://telecom.insa-lyon.fr/content/plateforme-smart-chappe-building}}. This plateform will allow us to demonstrate the integration and detail (i)~the interaction between smartphones and temperature/presence sensors in the building, (ii)~the interaction between smartphones and static Wifi hotspots and/or mobile service gateways and (iii)~the storage and computation migration of sensed data collected in the building to a cloud.

\bibliographystyle{abbrv}
\bibliography{Ubimob2011}

\end{document}